\IfFileExists{/dev/null}{\immediate\write18{sh makeLinks.sh}}{\immediate\write18{makeLinks.bat}}

\documentclass[a4paper,journal,twocolumn,10pt,final]{IEEEtran}

\makeatletter
\def\subsubsection{%
	\@startsection
	{subsubsection}                 
	{3}                             
	{\z@}                           
	{2.5ex plus 1.5ex minus 1.5ex}  
	{1ex plus .5ex minus 0ex}     
	{\normalfont\normalsize\itshape}
}
\makeatother

\usepackage[none]{hyphenat}
\usepackage[utf8]{inputenc}
\usepackage[T1]{fontenc}
\usepackage{textcomp}
\usepackage{scalefnt}
\usepackage{amsmath,amsfonts,amssymb,amsbsy,amstext}

\usepackage{mathtools}          
\usepackage{cmap}               
\usepackage[dvipsnames,svgnames,x11names]{xcolor}   
\usepackage{varwidth}

\let\originalleft\left
\let\originalright\right
\renewcommand{\left}{\mathopen{}\mathclose\bgroup\originalleft}
\renewcommand{\right}{\aftergroup\egroup\originalright}

\usepackage{graphicx}
\graphicspath{{.}{.\figs}}
\usepackage{placeins} 

\usepackage{multirow}
\usepackage{tabularx}
\newcolumntype{C}{>{\centering\arraybackslash}X}
\newcolumntype{R}{>{\flushright\arraybackslash}X}
\newcolumntype{L}{>{\flushleft\arraybackslash}X}
\newcolumntype{P}{>{\centering\arraybackslash} p{0.5\linewidth}}
\usepackage{booktabs}

\usepackage[%
style=ieee,
backend=biber, 
]{biblatex}

\defbibheading{bibliography}[\refname]{\section*{#1}}

\makeatletter
\g@addto@macro{\UrlBreaks}{\UrlOrds}
\makeatother

\usepackage{acro}[=v2]

\NewDocumentCommand{\acro}{m o m o}
{%
	\IfValueTF{#2}{%
		\IfValueTF{#4}{%
			\DeclareAcronym{#1}{short={#2},long={#3},#4}
		}{%
			\DeclareAcronym{#1}{short={#2},long={#3}}
		}
	}{%
		\IfValueTF{#4}{%
			\DeclareAcronym{#1}{short={#1},long={#3},#4}
		}{%
			\DeclareAcronym{#1}{short={#1},long={#3}}
		}
	}
}

\acro{2D}{two-dimensional}
\acro{3D}{three-dimensional}
\acro{4D}{four-dimensional}
\acro{6D}{six-dimensional}
\acro{1G}{first generation}
\acro{2G}{second generation}
\acro{3G}{third generation}
\acro{4G}{fourth generation}
\acro{5G}{fifth generation}
\acro{5GC}{5G core network}
\acro{5G-StoRM}{5G stochastic radio channel for dual mobility}
\acro{6G}{sixth generation}
\acro{3GPP}{3rd generation partnership project}
\acro{3GPP2}{3rd generation partnership project 2}
\acro{A2A}{air-to-air}
\acro{A2G}{air-to-ground}
\acro{AA}{antenna array}
\acro{AC}{admission control}
\acro{AD}{attack-decay}
\acro{AE}{antenna element}
\acro{AF}{amplify and forward}
\acro{ABS}{almost blank subframe}
\acro{ACF}{autocorrelation function}
\acro{ADSL}{asymmetric digital subscriber line}
\acro{AHW}{alternate hop-and-wait}
\acro{AI}{artificial intelligence}
\acro{AMC}{adaptive modulation and coding}
\acro{AP}{access point}
\acro{APA}{adaptive power allocation}
\acro{API}{application protocol interface}
\acro{ARQ}{automatic repeat request}
\acro{AT}{averaged time}
\acro{ARMA}{autoregressive moving average}
\acro{ASE}{average squared error}
\acro{ASC}{adaptive satisfaction control}
\acro{ATES}{adaptive throughput-based efficiency-satisfaction trade-off}
\acro{AWGN}{additive white gaussian noise}
\acro{B5G}{beyond 5G}
\acro{BAP}{backhaul adaptation protocol}
\acro{BB}{branch and bound}
\acro{BC}{branch and cut}
\acro{BD}{block diagonalization}
\acro{BER}{bit error rate}
\acro{BF}{best fit}
\acro{BL}{bit loading}
\acro{BLER}{block error rate}
\acro{BLPC-1}{bit loading with power constraint in hop 1}
\acro{BLPC-2}{bit loading with power constraint in hop 2}
\acro{BPC}{binary power control}
\acro{BPSK}{binary phase-shift keying}
\acro{BRA}{balanced random allocation}
\acro{BS}{base station}
\acro{BSP}{\acs*{BS} placement}
\acro{BSR}{buffer status report}
\acro{BoI}{band of interest}
\acro{C-link}{control link}
\acro{CAP}{combinatorial allocation problem}
\acro{CAPEX}{capital expenditure}
\acro{CB}{contextual bandit}
\acro{CBF}{coordinated beamforming}
\acro{CBR}{constant bit rate}
\acro{CBS}{class based scheduling}
\acro{CC}{congestion control}
\acro{CCL}{common cell list}
\acro{CDF}{cumulative distribution function}
\acro{CDL}{clustered delay line}
\acro{CDMA}{code-division multiple access}
\acro{CIR}{channel impulse response}
\acro{CH}{channel hardening}
\acro{CHO}{conditional handover}
\acro{C-RAN}{cloud-based radio access network}
\acro{CL}{closed loop}
\acro{CLT}{central limit theorem}
\acro{CLPC}{closed loop power control} 
\acro{CN}{core network}       
\acro{CNR}{channel-to-noise ratio}
\acro{CP}{control plane}
\acro{CPA}{cellular protection algorithm}
\acro{CPICH}{common pilot channel}
\acro{CoMP}{coordinated multi-point}
\acro{CQI}{channel quality indicator}
\acro{CRE}{cell range expansion}
\acro{CRM}{constrained rate maximization}
\acro{CRN}{cognitive radio network}
\acro{C-RNTI}{cell radio network temporary identifier}
\acro{CRRM}{centralized/common radio resource management}
\acro{CRS}{cell-specific reference signal}
\acro{CS}{coordinated scheduling}
\acro{CSI}{channel state information}
\acro{CSI-RS}{channel state information reference signal}
\acro{CTS}{clear to send}
\acro{CU}{centralized unit}
\acro{CUE}{cellular user equipment}
\acro{CWND}{congestion window size}
\acro{D2D}{device-to-device}
\acro{D2N}{drone-to-network}
\acro{DAA}{drone antenna array}
\acro{DAG}{directed acyclic graph}
\acro{DBS}{drone mounted base station}
\acro{DC}{dual connectivity}
\acro{DCA}{dynamic channel allocation}
\acro{DE}{differential evolution}
\acro{DF}{decode and forward}
\acro{DFT}{discrete fourier transform}
\acro{DIST}{distance}
\acro{DL}{downlink}
\acro{DMA}{double moving average}
\acro{DMRS}{demodulation reference signal}
\acro{D2DM}{\acs*{D2D} mode}
\acro{DMS}{\acs*{D2D} mode selection}
\acro{DP}{data plane}
\acro{DPC}{dirty paper coding}
\acro{DRA}{dynamic resource assignment}
\acro{DSA}{dynamic spectrum access}
\acro{DSM}{delay-based satisfaction maximization}
\acro{DU}{distributed unit}
\acro{E2E}{end-to-end}
\acro{ECC}{electronic communications committee}
\acro{EDF}{earliest deadline first}
\acro{EE}{energy efficiency}
\acro{EFLC}{error feedback based load control}
\acro{EI}{efficiency indicator}
\acro{e-ICIC}{enhanced inter-cell interference coordination}
\acro{eMBB}{enhanced mobile broadband}
\acro{EM}{electromagnetic}
\acro{eNB}{evolved node B}
\acro{EXP}{exponential}
\acro{EPA}{equal power allocation}
\acro{EPC}{evolved packet core}
\acro{EPS}{evolved packet system}
\acro{E-UTRAN}{evolved universal terrestrial radio access network}
\acro{ES}{exhaustive search}
\acro{FCP}{fundamental counting principle}
\acro{FCA}{flow control algorithm}
\acro{FD}{full duplex}
\acro{FDD}{frequency division duplex}
\acro{FDM}{frequency division multiplexing}
\acro{FDMA}{frequency division multiple access}
\acro{FER}{frame erasure rate}
\acro{FIFO}{first in first out}
\acro{FoV}{field-of-view}
\acro{FF}{fast fading}
\acro{FR}{frequency range}
\acro{FRS}[FS]{fast-rat scheduling}
\acro{FS}{fast switching}
\acro{FSB}{fixed switched beamforming}
\acro{FST}{fixed \acs*{SNR} target}
\acro{FT}{fourier transform}
\acro{FTP}{file transfer protocol}
\acro{Fwd}{forwarding}
\acro{GA}{genetic algorithm}
\acro{GAP}{generalized assignment problem}
\acro{GAP-MQ}{generalized assignment problem with minimum quantities}
\acro{GATES}{generalized adaptive throughput-based efficiency-satisfaction trade-off}
\acro{GBR}{guaranteed bit rate}
\acro{GLR}{gain to leakage ratio}
\acro{gNB}{gNodeB}
\acro{GOS}{generated orthogonal sequence}
\acro{GP}{gaussian process}
\acro{GPL}{GNU general public license}
\acro{GPS}{global positioning system}
\acro{GRP}{grouping}
\acro{GSM}{global system for mobile communications}
\acro{GTEL}{wireless telecommunications research group}
\acro{HARQ}{hybrid automatic repeat request}
\acro{HBF}{hybrid beamforming}
\acro{HCPP}{hardcore point process}
\acro{HetNet}{heterogeneous network}
\acro{HD}{half duplex}
\acro{HF}{high-frequency}
\acro{HH}{hughes-hartogs}
\acro{HardH}[HH]{hard handover}
\acro{HMS}{harmonic mode selection}
\acro{HO}{handover}
\acro{HOL}{head of line}
\acro{HPBW}{half power beamwidth}
\acro{HSDPA}{high speed downlink packet access}
\acro{HSR}{high-speed railway}
\acro{HSPA}{high speed packet access}
\acro{HTTP}{hypertext transfer protocol}
\acro{IAB}{integrated access and backhaul}
\acro{ICMP}{internet control message protocol} 
\acro{ICI}{intercell interference}
\acro{ICIC}{inter-cell interference coordination}
\acro{ID}{identification}
\acro{i.i.d.}{independent and identically distributed}
\acro{IETF}{internet engineering task force}
\acro{IPC}{individual power constraint}
\acro{UID}{unique identification}
\acro{IID}{independent and identically distributed}
\acro{IIR}{infinite impulse response}
\acro{ILP}{integer linear problem}
\acro{IMT}{international mobile telecommunications}
\acro{INV}{inverted norm-based grouping} 
\acro{IoT}{internet of things}
\acro{IP}{internet protocol}
\acro{IPv6}{internet protocol version 6}
\acro{IRA}{integrated resource allocation}
\acro{IRS}{intelligent reflecting surface}
\acro{ISD}{inter-site distance}
\acro{ISI}{inter symbol interference}
\acro{ISM}{industrial, scientific and medical}
\acro{ITU}{international telecommunication union}
\acro{JOAS}{joint opportunistic assignment and scheduling}
\acro{JOS}{joint opportunistic scheduling}
\acro{JP}{joint processing}
\acro{JRAPAP}{joint rb assignment and power allocation problem}
\acro{JS}{jump-stay}
\acro{JSM}{joint satisfaction maximization}
\acro{KKT}{karush-kuhn-tucker}
\acro{KPI}{key performance indicator}
\acro{LAC}{link admission control}
\acro{LA}{link adaptation}
\acro{LBS}{location based service}
\acro{LC}{load control}
\acro{LOS}{line of sight}
\acro{LP}{linear programming}
\acro{LTE}{long term evolution}
\acro{LTE-A}{\acs*{LTE}-advanced}
\acro{LTE-Advanced}{\ac{LTE-A}}
\acro{LTE-R}{\ac{LTE} railway}
\acro{LSP}{large scale parameter}
\acro{MADRDPG}{multi-agent deep recurrent deterministic policy gradient}
\acro{MeNB}{master \acs*{ENB}}
\acro{M2M}{machine-to-machine}
\acro{MAC}{medium access control}
\acro{MANET}{mobile ad hoc network}
\acro{MEDS}{method of exact doppler spread}
\acro{MC}{modular clock}
\acro{MCP}{minimal cost power}
\acro{MCS}{modulation and coding scheme}
\acro{MDB}{measured delay based}
\acro{MDI}{minimum \acs*{D2D} interference}
\acro{MDSM}{modified delay-based satisfaction maximization}
\acro{MDU}{max-delay-utility}
\acro{METIS}{mobile and wireless communications enablers for the twenty-twenty information society \acs*{5G}}
\acro{MF}{matched filter}
\acro{MFAP}{mobile femtocell access point}
\acro{MG}{maximum gain}
\acro{MH}{multi-hop}
\acro{mIAB}{mobile \ac{IAB}}
\acro{MILP}{mixed integer linear programming}
\acro{MIMO}{multiple input multiple output}
\acro{MINLP}{mixed integer nonlinear programming}
\acro{MIP}{mixed integer programming}
\acro{MISO}{multiple input single output}
\acro{MIT}{mobility interruption time}
\acro{ML}{machine learning}
\acro{MLWDF}{modified largest weighted delay first}
\acro{MME}{mobility management entity}
\acro{MMF}{max-min fairness}
\acro{mmMAGIC}{millimetre-wave based mobile radio access network for fifth generation integrated communications}
\acro{MMRP}{maximizing the minimal residual power}
\acro{MMRP-LB}{maximizing the minimal residual power with lower bound}
\acro{MMSE}{minimum mean square error}
\acro{mMTC}{massive machine-type communications}
\acro{mmWave}{millimeter wave}
\acro{MN}{master node}
\acro{MOS}{mean opinion score}
\acro{MPF}{multicarrier proportional fair}
\acro{MPRP}{maximization of the product of the residual powers}
\acro{MRA}{maximum rate allocation}
\acro{MR}{maximum rate}
\acro{MRC}{maximum ratio combining}
\acro{MRN}{mobile relay node}
\acro{MRT}{maximum ratio transmission}
\acro{MRUS}{maximum rate with user satisfaction}
\acro{MS}{mode selection}
\acro{MSE}{mean squared error}
\acro{MCG}{master cell group} 
\acro{MSI}{multi-stream interference}
\acro{MT}{mobile termination}
\acro{MTC}{machine-type communication}
\acro{IMS}{\acs*{IP} multimedia subsystem}
\acro{MTSI}{multimedia telephony services over \acs*{IMS}}
\acro{MTSM}{modified throughput-based satisfaction maximization}
\acro{MU-MIMO}{multi-user multiple input multiple output}
\acro{MU}{multi-user}
\acro{Multi-CUT}{multi-cell and multi-user and multi-tier}
\acro{NAS}{non-access stratum}
\acro{NB}{node B}
\acro{nBS}{neighbor base station}
\acro{NCL}{neighbor cell list}
\acro{NCR}{network-controlled repeater}
\acro{NG}{next generation}
\acro{NGC}{next generation core network}
\acro{NLP}{nonlinear programming}
\acro{NLOS}{non-line of sight}
\acro{NMSE}{normalized mean square error}
\acro{NN}{neural network}
\acro{NOMA}{non-orthogonal multiple access}
\acro{NORM}{normalized projection-based grouping}
\acro{NP}{non-polynomial time}
\acro{NR}{new radio}
\acro{NRT}{non-real time}
\acro{NSA}{non-stand-alone}
\acro{NSPS}{national security and public safety services}
\acro{OBF}{opportunistic beamforming}
\acro{OFDMA}{orthogonal frequency division multiple access}
\acro{OFDM}{orthogonal frequency division multiplexing}
\acro{OFPC}{open loop with fractional path loss compensation}
\acro{O2I}{outdoor-to-indoor}
\acro{OL}{open loop}
\acro{OLPC}{open-loop power control}
\acro{OL-PC}{open-loop power control}
\acro{OM}[O\&M]{operational \& maintenance}
\acro{OMA}{orthogonal multiple access}
\acro{OPEX}{operational expenditure}
\acro{ORA}{orthogonal resource allocation}
\acro{ORB}{orthogonal random beamforming}
\acro{JO-PF}{joint opportunistic proportional fair}
\acro{OSI}{open systems interconnection}
\acro{PA}{power allocation}
\acro{PAIR}{\acs*{D2D} pair gain-based grouping}
\acro{PAPR}{peak-to-average power ratio}
\acro{P2P}{peer-to-peer}
\acro{PBCH}{physical broadcast channel}
\acro{PBS}{pico base station}        
\acro{PC}{power control}
\acro{PCI}{physical cell \acs*{ID}}
\acro{PDCP}{packet data convergence protocol}
\acro{PDF}{probability density function}
\acro{PDU}{protocol data unit}
\acro{PER}{packet error rate}
\acro{PF}{proportional fair}
\acro{PGRA}{probabilistic graph based resource allocation}
\acro{P-GW}{packet data network gateway}
\acro{PHY}{physical}
\acro{PL}{pathloss}
\acro{PLR}{packet loss ratio}
\acro{PRABE}{power and resource allocation based on quality of experience}
\acro{PRB}{physical resource block}
\acro{PROJ}{projection-based grouping}
\acro{PSD}{power spectral density}
\acro{PSDe}{positive semi-definite}
\acro{ProSe}{proximity services}
\acro{PS}{packet scheduling}
\acro{PSO}{particle swarm optimization}
\acro{PSS}{primary synchronization signal}
\acro{PTAS}{polynomial-time approximation scheme}
\acro{PTP}{point-to-point}
\acro{PZF}{projected zero-forcing}
\acro{QAM}{quadrature amplitude modulation}
\acro{QHMLWDF}{queue-HOL-MLWDF}
\acro{QoE}{quality of experience}
\acro{QoS}{quality of service}
\acro{QPSK}{quadri-phase shift keying}
\acro{QSM}{queue-based satisfaction maximization}
\acro{QuaDRiGa}{quasi deterministic radio channel generator}
\acro{RACH}{random access}
\acro{RAISES}{reallocation-based assignment for improved spectral efficiency and satisfaction}
\acro{RAN}{radio access network}
\acro{RA}{resource allocation}
\acro{RAP}{rb assignment problem}
\acro{RAT}{radio access technology}[long-plural-form={radio access technologies}]
\acro{RATE}{rate-based}
\acro{RAU}{remote antenna unit}
\acro{RB}{resource block}
\acro{RBG}{resource block group}
\acro{REF}{reference grouping}
\acro{RET}{remote electrical tilt}
\acro{RF}{radio frequency}
\acro{RL}{reinforcement learning}
\acro{RLC}{radio link control}
\acro{RLF}{radio link failure}
\acro{RM}{rate maximization}
\acro{RMa}{rural macro}
\acro{RMJ-SNR}{region of the minimum joint snr}
\acro{RMEC}{rate maximization under experience constraints}
\acro{RMSE}{root-mean-square error}
\acro{RNC}{radio network controller}
\acro{RND}{random grouping}
\acro{RNN}{recurrent neural network}
\acro{RRA}{radio resource allocation}
\acro{RRM}{radio resource management}
\acro{RSCP}{received signal code power}
\acro{RSRP}{reference signal received power}
\acro{RSRQ}{reference signal received quality}
\acro{RR}{round robin}
\acro{RIS}{reconfigurable intelligent surface}
\acro{RRC}{radio resource control}
\acro{RSSI}{received signal strength indicator}
\acro{RT}{real time}
\acro{RTS}{request to send}
\acro{RU}{resource unit}
\acro{RUNE}{rudimentary network emulator}
\acro{RV}{random variable}
\acro{Rx}{receiver}
\acro{RZF}{regularized zero-forcing}
\acro{SA}{subcarrier assignment}
\acro{SAC}{session admission control}
\acro{sBS}{serving base station}
\acro{SC}{small cell}
\acro{SCI}{side control information}
\acro{SCon}[SC]{single connectivity}
\acro{SCG}{secondary cell group}
\acro{SCM}{spatial channel model}
\acro{SCS}{subcarrier spacing}
\acro{SC-FDMA}{single carrier - frequency division multiple access}
\acro{SCRV}{spatially consistent random variable}
\acro{SD}{soft dropping}
\acro{SDAP}{service	data adaptation protocol}
\acro{S-D}{source-destination}
\acro{SDPC}{soft dropping power control}
\acro{SDM}{space-division multiplexing}
\acro{SDMA}{space-division multiple access}
\acro{SE}{squared error}
\acro{SF}{shadow fading}
\acro{SMDP}{semi-markov	decision problem}
\acro{SeNB}{secondary \acs*{ENB}}
\acro{SER}{symbol error rate}
\acro{SES}{simple exponential smoothing}
\acro{S-GW}{serving gateway}
\acro{SIC}{successive interference cancellation}
\acro{SelfIC}[SIC]{self interference cancellation}
\acro{SINR}{signal to interference-plus-noise ratio}
\acro{SLNR}{signal to leakage-plus-noise ratio}
\acro{SNN}{strictly non-negative}
\acro{SI}{satisfaction indicator}
\acro{SelfI}[SI]{self interference}
\acro{SIP}{session initiation protocol}
\acro{SISO}{single input single output}
\acro{SIMO}{single input multiple output}
\acro{SIR}{signal to interference ratio}
\acro{SM}{subcarrier matching}
\acro{SMA}{simple moving average}
\acro{SMSE}{spatial-mean-square error}
\acro{SN}{secondary node}
\acro{SNR}{signal to noise ratio}
\acro{SON}{self organizing networks}
\acro{SoA}{state-of-the-art}
\acro{SoS}{sum-of-sinusoids}
\acro{SORA}{satisfaction oriented resource allocation}
\acro{SORA-NRT}{satisfaction-oriented resource allocation for non-real time services}
\acro{SORA-RT}{satisfaction-oriented resource allocation for real time services}
\acro{SP}{signal processing}
\acro{SPF}{single-carrier proportional fair}
\acro{SR}{smart repeater}
\acro{ASR}{advanced \ac{SR}}
\acro{SRA}{sequential removal algorithm}
\acro{SRB1}{signaling radio bearer~1}
\acro{SRS}{sounding reference signal}
\acro{SSB}{synchronization signal block}
\acro{SSP}{small scale parameter}
\acro{SSS}{secondary synchronization signal}
\acro{ST}{spanning tree}
\acro{STTD}{space time transmit diversity}[long-plural-form={space time transmit diversities}]
\acro{SU-MIMO}{single-user multiple input multiple output}
\acro{SU}{single-user}
\acro{tBS}{target base station}
\acro{SVD}{singular value decomposition}
\acro{TCP}{transmission control protocol}
\acro{TDD}{time division duplex}
\acro{TDM}{time division multiplexing}
\acro{TDMA}{time division multiple access}
\acro{TDMed}{time division multiplexed}
\acro{TETRA}{terrestrial trunked radio}
\acro{TP}{transmit power}
\acro{TPC}{total power constraint}
\acro{TR}{technical report}
\acro{TTI}{transmission time interval}
\acro{TTR}{time-to-rendezvous}
\acro{TTT}{time-to-trigger}
\acro{TSM}{throughput-based satisfaction maximization}
\acro{TU}{typical urban}
\acro{TV}{television}
\acro{TVWS}{\acs*{TV} white space}
\acro{Tx}{transmitter}
\acro{UAV}{unmanned aerial vehicle}
\acro{UABSC}{user-assisted bearer split control}
\acro{UDP}{user datagram protocol}
\acro{UDN}{ultra-dense networks}
\acro{UE}{user equipment}
\acro{UBA}{\ac{UE}-\ac{BS} association}
\acro{UEPS}{urgency and efficiency-based packet scheduling}
\acro{UFC}{federal university of cear\'{a}}
\acro{ULA}{uniform linear array}
\acro{UL}{uplink}
\acro{UMa}{urban macro}
\acro{UMi}{urban micro}
\acro{UMTS}{universal mobile telecommunications system}
\acro{UP}{user plane}
\acro{UPN}{user provided network}
\acro{URA}{uniform rectangular array}
\acro{URI}{uniform resource identifier}
\acro{URLLC}{ultra-reliable low-latency communications}
\acro{URM}{unconstrained rate maximization}
\acro{VBR}{variable bit rate}
\acro{VET}{variable electrical tilt}
\acro{VMR}{vehicle-mounted relay}
\acro{VR}{virtual resource}
\acro{VoIP}{voice over \acs*{IP}}
\acro{WCDMA}{wideband code division multiple access}
\acro{WF}{water-filling}
\acro{WID}{wireless infrastructure drone}
\acro{Wi-Fi}{wireless fidelity}
\acro{WiMAX}{worldwide interoperability for microwave access}
\acro{WINNER}{wireless world initiative new radio}
\acro{WLAN}{wireless local area network}
\acro{WMPF}{weighted multicarrier proportional fair}
\acro{WP}{work package}
\acro{WPF}{weighted proportional fair}
\acro{WSN}{wireless sensor network}
\acro{WWW}{world wide web}
\acro{WFQ}{weighted fair queuing}
\acro{XIXO}{(single or multiple) input (single or multiple) output}
\acro{XPR}{cross-polarization power ratio}
\acro{ZF}{zero-forcing}
\acro{ZMCSCG}{zero mean circularly symmetric complex gaussian}

\usepackage{siunitx}
\usepackage{datetime}
\usepackage{url}

\usepackage[caption=false,font=footnotesize]{subfig}

\usepackage{algorithm}
\usepackage{algorithmicx}
\usepackage{algpseudocode}

\usepackage{epstopdf}

\usepackage[hidelinks=true]{hyperref}

\usepackage{enumerate}

\usepackage[final]{changes} 
\definechangesauthor[name={Victor F. Monteiro}]{vfm}
\definechangesauthor[name={Tarcisio F. Maciel}]{tfm}
\setaddedmarkup{{\color{blue}#1}}
\setdeletedmarkup{{\color{red}\sout{#1}}}

\usepackage[referable,flushleft]{threeparttablex}
\AtBeginEnvironment{tablenotes}{\footnotesize} 

\DeclareMathAlphabet{\mathppl}{T1}{ppl}{m}{it}
\DeclareMathAlphabet{\mathphv}{T1}{phv}{m}{it}
\DeclareMathAlphabet{\mathpzc}{T1}{pzc}{m}{it}

\newcommand{\FigRef}[2][]{Fig.#1~\ref{#2}}

\newcommand{\TabRef}[2][]{Table#1~\ref{#2}}

\usepackage{standalone}
\usepackage{shapepar}

\usepackage{tikz}
\usetikzlibrary{arrows}
\usetikzlibrary{positioning}
\usetikzlibrary{shapes.misc}
\usetikzlibrary{shapes.geometric}
\usetikzlibrary{shapes.symbols}
\usetikzlibrary{external}

\usetikzlibrary{shadows}
\usetikzlibrary{backgrounds}
\usetikzlibrary{shapes}
\usetikzlibrary{shapes.multipart}
\usetikzlibrary{matrix}
\usetikzlibrary{intersections}
\usetikzlibrary{fit}
\usetikzlibrary{calc}
\usetikzlibrary{chains}
\usetikzlibrary{scopes}
\usetikzlibrary{decorations.pathreplacing}
\usetikzlibrary{decorations.text}
\usetikzlibrary{arrows.meta} 
\usetikzlibrary{mindmap}
\usetikzlibrary{trees}

\usepackage{pgfplots}
\pgfplotsset{%
	width=0.95\columnwidth,
	height=0.25\textheight, 
	compat=1.14,
	compat/show suggested version=false,
	filter discard warning=false,
	tick label style={font=\footnotesize},
	label style={font=\footnotesize},
	every axis label={font=\footnotesize},
	grid=major,
	grid style={dashed,gray!30},
	cycle list shift=0,
	enlargelimits=false,
	legend style={%
		font=\footnotesize,
		legend cell align=left,
		nodes={inner xsep=2pt,inner ysep=1pt,text depth=0.15em},
	},
}
\usepackage{pgfplotstable}

\newtoggle{tikzexternalize}
\togglefalse{tikzexternalize}

\iftoggle{tikzexternalize}{
	\immediate\write18{mkdir -p ./tikz-external-figs/}
	\usetikzlibrary{external}
	\tikzexternalize[prefix=./tikz-external-figs/,mode=list and make]
	\tikzset{external/system call={pdflatex \tikzexternalcheckshellescape -halt-on-error -interaction=batchmode -jobname "\image" "\texsource"}}
}{
}

\AtBeginDocument{%
	\abovedisplayskip 1.5ex plus 4pt minus 2pt
	\belowdisplayskip \abovedisplayskip
	\abovedisplayshortskip 0pt plus 4pt
	\belowdisplayshortskip 1.5ex plus 4pt minus 2pt
}

\usepackage{tikz}
\usetikzlibrary{arrows}
\usetikzlibrary{positioning}

\usetikzlibrary{shapes}
\usetikzlibrary{matrix}
\usetikzlibrary{intersections}
\usetikzlibrary{fit}
\usetikzlibrary{calc}
\usetikzlibrary{chains}

\usepackage{pgfplots}

\definecolor{colorBlue}{RGB}{31,119,180}
\definecolor{colorOrange}{RGB}{255,127,14}
\definecolor{colorGreen}{RGB}{67,170,67}
\definecolor{colorDarkGreen}{RGB}{0,128,0}

\pgfplotsset{common line style/.style={line width=1pt}}

\pgfplotsset{every axis plot post/.append style={
    every mark/.append style={scale=1.5}
}}

\pgfplotsset{common plots axis options/.style={
	every axis/.append style={
	  legend style={fill=gray!5, fill opacity=0.85, text opacity=1}
	},
	width=\columnwidth,
	height=0.4\columnwidth,
	grid=both,
	filter discard warning=false,
	tick label style={font=\footnotesize},
	label style={font=\footnotesize},
	every axis label={font=\footnotesize},
	grid=major,
	grid style={dashed,gray!30},
	cycle list shift=0,
	enlargelimits={true,abs value=1pt},
	ylabel shift = -0.5ex,
	legend style={%
		font=\scriptsize,
		legend cell align=left,
		nodes={inner xsep=2pt,inner ysep=1pt,text depth=0.15em},
	},
	}
}

\pgfplotsset{bar axis options/.style={
		common plots axis options,
		ybar=1pt,
		bar width = 3pt,
		enlarge x limits={true,abs value=5pt},
}}

\pgfplotsset{mcs axis options/.style={
		bar axis options,
		width=\columnwidth,
		height=0.4\columnwidth,
		ymin = 0, ymax = 100,
		xtick={0, 1, ..., 15},
		xticklabels={Total, MCS 1, MCS 2, MCS 3, MCS 4, MCS 5, MCS 6, MCS 7, MCS 8, MCS 9, MCS 10, MCS 11, MCS 12, MCS 13, MCS 14, MCS 15},
		x tick label style={
			font=\tiny,
			xshift = 1ex,
			rotate=45,
			anchor=east,
		},
}}

\pgfplotsset{common marker style/.style={
		mark repeat = 10,
		mark size = 1pt,
		mark options={solid},
}}

\pgfplotsset{scenario311Dir style/.style={
	common line style,
	common marker style,
	MediumSeaGreen,
	mark=x,
	mark size=1.3pt,
}}

\pgfplotsset{scenario322Dir style/.style={
	common line style,	
	common marker style,
	Red,
	mark=*,
	mark size=0.8pt,
}}

\pgfplotsset{scenario01Dir style/.style={
    common line style,
    common marker style,
    DodgerBlue,
    mark=triangle*,
    mark size=1pt,
}}

\pgfplotsset{scenario02Dir style/.style={
		common line style,
		common marker style,
		Orange,
		mark=square*,
		mark size=0.7pt,
}}

\pgfplotsset{scenario311Ncr style/.style={
		dashed,
		line width=1.1pt,
		common marker style,
		MediumSeaGreen,
		mark=x,
		mark size=1.0pt,
}}

\pgfplotsset{scenario322Ncr style/.style={
		dashed,
		line width=1.1pt,
		common marker style,
		Red,
		mark=*,
		mark size=0.8pt,
}}

\pgfplotsset{scenario01Ncr style/.style={
		dashed,
		line width=1.1pt,
		common marker style,
		DodgerBlue,
		mark=triangle*,
		mark size=1.0pt,
}}

\pgfplotsset{scenario02Ncr style/.style={
		dashed,
		line width=1.1pt,
		common marker style,
		Orange,
		mark=square*,
		mark size=0.8pt,
}}

\pgfplotsset{load bar style/.style={
		DodgerBlue, fill
}}

\pgfplotsset{ack bar style/.style={
		DodgerBlue, fill
}}

\pgfplotsset{nack bar style/.style={
		Red, fill
}}

\def\plotsDataPath{figs/plots/data}

\begin{document}
\title{Network-Controlled Repeater - An Introduction}

\author{Fco. Italo G. Carvalho, Raul Victor de O. Paiva, Tarcisio F. Maciel,~\IEEEmembership{Senior Member,~IEEE}, \\  Victor F. Monteiro,~\IEEEmembership{Member,~IEEE}, Fco. Rafael M. Lima,~\IEEEmembership{Senior Member,~IEEE}, Darlan C. Moreira, \\ Diego A. Sousa, Behrooz Makki,~\IEEEmembership{Senior Member,~IEEE}, Magnus Åstr$\ddot{\text{o}}$m and Lei Bao
}

\maketitle

\begin{abstract}
In \ac{5G} wireless cellular networks, millimeter wave spectrum opens room for several potential improvements in throughput, reliability, latency, among other aspects. %
However, it also brings challenges, such as a higher influence of blockage which may significantly limit the coverage. %
In this context, \acp{NCR} are network nodes with low complexity that represent a technique to overcome coverage problems. %
In this paper, we introduce the \ac{NCR} concept and study its performance gains and deployment options. %
Particularly, presenting the main specifications of \acs{NCR} as agreed in \ac{3GPP} Rel-18, we analyze different \ac{NCR} deployments in an urban scenario and compare its performance with alternative deployments. %
As demonstrated, with a proper network planning  and beamforming design, \ac{NCR} is an attractive solution to cover blind spots the \acp{BS} may have. %
\end{abstract}

\begin{IEEEkeywords}
	Network-controlled repeater, 3GPP, network densification, 5G, beamforming, millimeter wave communications
\end{IEEEkeywords}

%
\IEEEpeerreviewmaketitle
\acresetall

\section{Introduction} \label{SEC:Intro}

\Ac{5G} and beyond aim to provide \emph{everyone everywhere} with high \ac{QoS}. %
To cope with such requirements, different techniques are used, among which network densification, beamforming and \ac{mmWave} communications. 
Particularly, it is expected that, in near future, to assist the \acp{gNB}, a wireless network will be densified with different types of beamforming-capable nodes such as \ac{IAB} nodes, \acp{NCR} (sometimes called NetCRs), etc. %

During the \ac{3GPP} Rel- 16-18, \ac{IAB} has been specified as the main \ac{5G} relaying technique, which is based on decode-and-forwarding~\cite{3gpp.38.174c, Monteiro2022, Madapatha2020}.  %
In a few words, the goal of \acs{IAB} is to provide flexible wireless backhaul using \acs{3GPP} \ac{NR} technology, providing not only backhaul but also the existing cellular services in the same node. %
Although the coverage of \ac{IAB} is typically larger than the coverage of, e.g., Rel-17 and Rel-18 repeaters, it is a relatively complex and expensive node compared to other relaying nodes. %
Thus, depending on the deployment, there may be need for simpler and more affordable nodes with low complexity for, e.g., \textit{coverage hole} removal. %
Indeed, Rel-17 repeater~\cite{3gpp.38.106} is one option, which simply amplifies-and-forwards every signal it receives. %

Historically, repeaters are considered as a source of interference (since they forward everything including noise). %
So, it is beneficial to limit their interference via, e.g., proper beamforming. %
Additionally, as explained in the sequel, enabling the \acp{gNB} to control the repeaters makes it possible to deploy them in \ac{mmWave} spectrum where beamforming is a necessity, improve their data transmission and/or energy efficiency. %
These have been the main motivations for \ac{3GPP} to define and specify the requirements/capabilities for \ac{NCR} in Rel-18, as repeaters with beamforming capabilities which are under network control. %

In this paper, we introduce the \ac{NCR}, as defined by \ac{3GPP} Rel-18, and study their potential gains/deployment options. %
Highlighting the main Rel-18 \ac{NCR}-specific agreements on different protocol layers, we analyze different \ac{NCR} deployments in an urban scenario. %
Furthermore, we compare the system-level performance of \ac{NCR}-assisted networks with those in alternative network deployments. %
As we show, \acp{NCR} are attractive nodes to assist existing networks, especially for covering blind spots that are naturally inherent in present \ac{mmWave} networks. %
Additionally, network planning and beamforming are shown to have large impact on the performance of \ac{NCR}-assisted networks. %

It is worth noting that the most relevant works to our paper are~\cite{Flamini2022, Leone2022, Guo2022}, in which the concept of a heterogeneous smart electromagnetic environment with different types of network nodes is studied~\cite{Flamini2022}, network planning is optimized for the cases with \acp{NCR} and \acp{RIS}~\cite{Leone2022} and conceptual comparisons between \acp{NCR} and \acp{RIS} are presented~\cite{Guo2022}.  %

\section{\acs{3GPP} \acl{NCR}} \label{SEC:NCR}

In general, \acp{NCR} are repeaters with beamforming capability that can receive and process \ac{SCI} from the network. %
\acp{NCR} are of interest in both outdoor and indoor networks as well as for outdoor-to-indoor or indoor-to-outdoor communications. %
Different from \ac{IAB} which is based on decode-and-forward relaying, \acp{NCR} are beamforming-capable amplify-and-forward relay nodes and, thereby, amplify the noise/interferences. %
 The \ac{SCI} allows the \acp{NCR} to perform their amplify-and-forward operation efficiently. %
In other words, \acp{NCR} can be considered as network-controlled ``beam benders'' relative to the \ac{gNB}. %
That is, although deployed at a distance from their controlling \ac{gNB}, they are logically part of it for all management purposes, i.e., they are deployed under the control of the operator. %
Particularly, \ac{3GPP} has specified \acp{NCR} based on the following assumptions~\cite{3gpp.RP-222673}: %
\begin{itemize}
	\item \acp{NCR} are in-band \ac{RF} repeaters used for extension of network coverage on \ac{FR}~1\footnote{FR1 is defined as bands in the sub-6 GHz spectrum (although 7125 MHz is the maximum) while FR2 defines bands in the \ac{mmWave} range.} and \ac{FR}~2 bands based on the \ac{NCR} model given in~\cite{3gpp.38.867};
	\item \acp{NCR} are deployed in single hop stationary deployments;
	\item \acp{NCR} are transparent to the \ac{UE}, in the sense that the \ac{UE} is not made aware of the \acp{NCR} by the network, following its normal behavior as in the cases without \acp{NCR}; 
	\item \acp{NCR} can maintain the \ac{gNB}-repeater link and the repeater-\ac{UE} link simultaneously.
\end{itemize}

\begin{figure}[t]
	\centering
	\includegraphics[width=0.9\columnwidth]{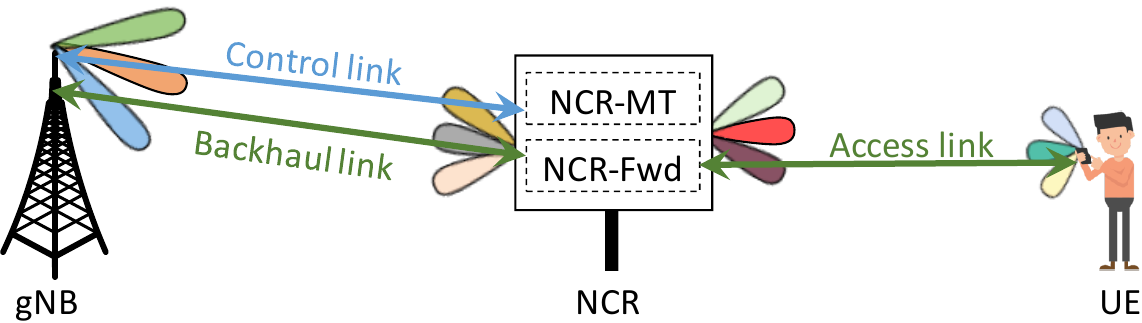}
	\caption{\acs{gNB} to \acs{UE} connection via an \ac{NCR}.}
	\label{FIG:link-selection}
\end{figure}

Figure~\ref{FIG:link-selection} demonstrates the schematic of an \ac{NCR} as defined in~\cite{3gpp.RP-222673}. %
The \ac{NCR} consists of two functions, namely, \ac{MT} and \ac{Fwd}. %
The \ac{NCR}-\ac{MT} is the entity responsible for communicating with its controlling \ac{gNB} to exchange information, e.g., \ac{SCI}. %
The link between the \ac{gNB} and the \ac{NCR}-\ac{MT} is referred to as \ac{C-link} which is based on \acs{NR} Uu interface. %
The \ac{NCR}-\ac{Fwd} is the entity that performs amplify-and-forward in the \ac{UL}/\ac{DL} signals between the \ac{gNB} and the \ac{UE}. %
The links between the \ac{gNB} and \ac{NCR}-\ac{Fwd} and between the \ac{NCR}-\ac{Fwd} and \ac{UE} are referred to as backhaul and access links, respectively. %
The behavior of the \ac{NCR}-\ac{Fwd} is solely controlled based on the \ac{SCI} received by the \ac{NCR}-\ac{MT} from the controlling \ac{gNB}. %
The \ac{3GPP} Rel-18 \ac{NCR} work-item has mainly focused on specifying the signaling of the \ac{SCI} for controlling the \ac{NCR}-\ac{Fwd} behavior related to beamforming, \ac{UL}-\ac{DL} \ac{TDD} operation and ON-OFF configuration. %

As one of its key properties, an \ac{NCR} can be turned ON/OFF to save energy/manage interference. %
Here, by default an \ac{NCR} is OFF, in which case the \ac{NCR}-\ac{Fwd} is not expected to forward signals. %
An \ac{NCR} goes into ON state only in the time resources with an access link beam indication. %
Both slot- and symbol-level granularities are feasible for the time-domain resource indication and determination of the access link beam. %


The \ac{NCR} receives a signal from the \ac{gNB} in \ac{DL} (or, from the \ac{UE} in \ac{UL}). %
Within the \ac{NCR}, the signal is amplified and possibly also filtered before it is transmitted to the \ac{UE} in \ac{DL} (or, to the \ac{gNB} in \ac{UL}). %
Beamforming is used for both reception and transmission, resulting in improved \ac{SINR}. %
Which beamforming to apply and when to apply it is provided to the \ac{NCR} over the \ac{SCI}. %
Consequently, the \ac{NCR} is not required to have any signal and channel awareness for signals forwarded by the \ac{NCR}-\ac{Fwd}. %
The forwarding operation is performed entirely in the analog domain, implying a latency in the order of tens of nanoseconds, and it is fully manageable (if at all needed) with existing timing alignment procedures. %
To receive SCI, the \ac{NCR}-\ac{MT} uses the normal Uu interface and must therefore support a subset of UE capabilities. %

Like fixed \ac{IAB}-nodes, \acp{NCR} are stationary nodes deployed according to the network planning done by the operator. %
The planning can be expected to result in a stable control/backhaul link between the controlling \ac{gNB} and \ac{NCR}.  %
Thus, the required features/capabilities of \ac{NCR}-\ac{MT} are like those in \ac{IAB}-\ac{MT} (see~\cite{3gpp.38.174c} and \cite{3gpp.38.306} for detailed specifications for \ac{IAB}). %

As baseline, the same large-scale properties of the channel are expected to be experienced by the \ac{C-link} and backhaul link, at least when the \ac{NCR}-\ac{MT} and \ac{NCR}-\ac{Fwd} operate in the same carrier as in Rel-18. %
The \ac{DL} communication of \ac{C-link} and backhaul link can be simultaneously performed or \ac{TDMed}. %
The \ac{UL} communication of \ac{C-link} and backhaul link, on the other hand, can be performed at least \ac{TDMed}. %
The simultaneous \ac{UL} transmission of the \ac{C-link} and backhaul link is subject to \ac{NCR} capability. %

Fixed and adaptive beams can be considered at \ac{NCR} for C-link and backhaul-link, since support for a fixed beam is part of the basic \ac{NCR} features, and, as a capability, the \ac{NCR}-\ac{MT} may perform adaptive beamforming on the C-link/backhaul link. %
On the one hand, in simultaneous operation, \ac{NCR}-Fwd follows \ac{NCR}-MT's beam indication (TCI or SRI). %
On the other hand, in non-simultaneous operation, \ac{NCR}-Fwd follows dedicated backhaul beam indication or pre-defined rules. %
Conditioned on capability, the backhaul link may additionally be configured with separate TCI states. %

The beam control of the \ac{NCR} in the access link should be carefully addressed to minimize the impact on cell-common and \ac{UE}-specific signals/channels which are forwarded towards the \acp{UE}. %
According to the specification, the forward resource can be configured periodically, semi-persistent, or dynamically. %
Periodic and semi-persistent beam configuration can be used, e.g., for repeating cell-specific signals, with rarely/never changed configuration and are configured per periodicity, where each periodicity includes a set of time offsets and durations. %
Additionally, dynamic access link beam configuration may be used for forwarding dynamically scheduled \ac{UE}-specific channels, which are indicated as pairs of beam and time resource indices, where the full set of time resources is preconfigured by \ac{RRC}.

The \ac{DL}/\ac{UL} transmission direction of the \ac{NCR}-\ac{Fwd} function can be principally derived from the \ac{NCR}-\ac{MT} configurations. %
Finally, \acp{NCR} do not support dynamic \ac{TDD} since that would require signal awareness and a significantly more complicated \ac{NCR} design. %

Different ranges of \acp{NCR}' links have been defined. %
At the \ac{UE}-side of \ac{NCR}, wide-, medium- and local-area range have been defined. %
At the \ac{gNB}-side, on the other hand, wide- and local-area range have been defined. %
These types of links’ properties can be independently declared. %
The main differences between wide- and local-area links are the level of network planning and transmit power. %
Wide-area links benefit from high transmit power/amplification gain and are well-planned where, for instance, the \ac{gNB}-side of the \ac{NCR} is located on the rooftop or similar, to achieve \ac{LOS} over a long distance to a controlling \ac{gNB}. %
For the local-area type, on the other hand, the node may be mounted at low heights, e.g., lamppost, etc., and may have a power in the order of a typical \ac{UE}. %

In the initial discussions to define the Rel-18 \ac{NCR} study-item, there have been discussions whether \acp{RIS} should also be included into the \ac{NCR} discussions. %
However, it was decided not to include and leave them for possible future releases. %
There are similarities and differences between \acp{NCR} and \acp{RIS}. %
In simple words, \acp{RIS} are \acp{NCR} with negative amplification gain, and both emit the incoming signals without decoding. %
Compared to \acp{NCR}, \acp{RIS} are expected to be simpler nodes with less focused beamforming capability/accuracy and without active amplification (or small amplification in the cases with active \ac{RIS}). %
As opposed to \acp{NCR}, \acp{RIS} do not amplify the signal but also not the noise/interference. %
Additionally, \acp{RIS} have only one reflection matrix, while \acp{NCR} can do separate beamforming at the transmitter and receiver sides. %
This should be an advantage for the \acp{NCR} when controlling the interference in, e.g., multi-user scenarios. %

\section{Performance Evaluation of \acs{NCR}-assisted Networks}
\label{NCR_LOS_SEC:Sys_Mod}

To study the performance of \acp{NCR}, we consider both \ac{UL} and \ac{DL} of a \ac{TDD} \ac{5G} \ac{NR} network, where \acp{NCR} are used to extend coverage and overcome possible coverage holes due to blockage. %
We assume the same type of data traffic in both link directions. 

Consider a network composed of $N$~\acp{NCR} and $B$~\acp{gNB} serving $U$~\acp{UE}, either directly or through one of the \acp{NCR}, over $K$~\acp{RB}, which correspond to the minimum allocable time-frequency system resources. %
At each allocation period, namely \ac{TTI}, each \ac{gNB} assigns \acp{RB} to its served \acp{UE} orthogonally, either directly or via an \ac{NCR}. %
Thus, the links of the same \ac{gNB} do not interfere with each other but may interfere with links from other \acp{gNB} reusing the same \acp{RB}. %
It is assumed that \acp{NCR} do not perform signal processing, i.e., do not filter out incoming signals that are transmitted to \acp{UE} who are not served by them.  
In other words, \acp{NCR} amplify and forward both desired and interfering signals. %
The \acp{gNB} and \acp{NCR} are equipped with antenna arrays while \acp{UE} have a single antenna. %

Considering \ac{DL} communication, the signal directly received by a \ac{UE} from its serving \ac{gNB} or indirectly via the \ac{gNB}-controlled \ac{NCR} might be formed by the superposition of several signal components, i.e.,
\begin{enumerate}[\noindent i)]
\item signal of interest coming from the serving \ac{gNB}, in case of direct communication, and/or an amplified and forwarded signal of interest coming from the \ac{NCR} serving the \ac{UE}; \label{item_comp_s}%
\item amplified \ac{AWGN} of the receiver of the serving \ac{NCR}, which is forwarded to the \ac{UE} in case of indirect communication; \label{item_comp_w1}%
\item \ac{AWGN} of the \ac{UE} receiver; \label{item_comp_w2} %
\item interfering signal components coming directly from neighbor \acp{gNB}; \label{item_comp_i1} %
\item interfering signals amplified and forwarded by neighbor \acp{NCR}; and \label{item_comp_i2} %
\item amplified \ac{AWGN} components from neighbor \acp{NCR} receivers, which are forwarded towards the \ac{UE}. \label{item_comp_w3}%
\end{enumerate} %
These components are subject to effects impaired by wireless channels, e.g., path loss, shadowing, fast fading, etc. %
They are also subjected to the gains of transmit/receive filters adopted by the \acp{gNB} and \acp{NCR}. %
A similar set of signals can be considered in \ac{UL} communication. %

In the considered scenario, the \ac{UE} association to either \acp{gNB} or \acp{NCR} is performed as follows. %
Beam sweepings between \ac{gNB} and \ac{NCR} and between \ac{NCR} and \ac{UE} are periodically performed with different periodicities, selected considering the profile of each type of link~\cite{3gpp.38.331b}. %
Then, the best beam pairs for each link (backhaul and/or access) are selected and used until the next beam sweeping
period. %
In this context, \acp{gNB} and \acp{NCR} are static nodes with \ac{LOS} established during the network planning phase. %
Then, the \ac{gNB}-\ac{NCR} link usually has a long coherence time and, therefore, requires less beam update (or even never requires). %
Oppositely, the beamforming at the access link may require frequent updates since the channel varies faster due to the \acp{UE}' mobility~\cite{3gpp.38.331b}. %
The \ac{UE} follows its typical behavior to measure the \ac{RSRP} of the received reference signals and reports the measurements to its serving \ac{gNB}. %
Based on the received \ac{RSRP} measurement reports, the \ac{gNB} evaluates whether to directly transmit to the \ac{UE} or to serve the UE through its \acp{NCR}. %

A key objective of \acp{NCR} is to extend/improve the coverage of \ac{5G} networks which tends to be less consistent at \ac{mmWave} bands. %
Herein, we aim to contribute to the research on \acp{NCR} by assessing the gains of introducing different numbers of
\acp{NCR} into networks covered originally only by conventional \acp{gNB}. %
\begin{figure}[t]
  \centering
  \includegraphics[width=0.98\columnwidth]{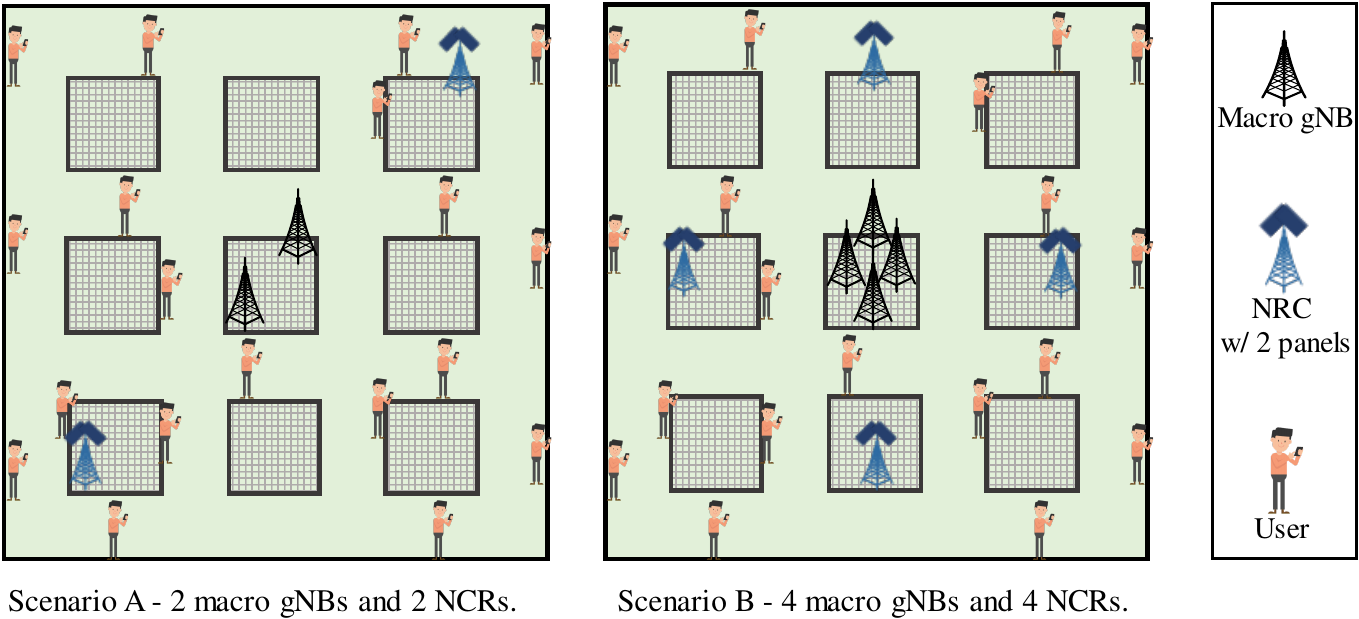}
  \caption{Scenarios of interest.}
  \label{NCR_LOS_FIG:scenario}
\end{figure} %
For this purpose, we analyze the system performance of different setups. %
Two base scenarios, as shown in~\FigRef{NCR_LOS_FIG:scenario}, are considered. %
In Scenario~A, we consider two macro-\acp{gNB} at the central block and two \acp{NCR}. %
Scenario~A is split into two cases: one case in which the shown \acp{gNB} and \acp{NCR} are present and another case where only the \acp{gNB} are used, which we refer to as macro-only scenario. %
Similarly, Scenario~B contains four macro-\acp{gNB} at the central block, and four \acp{NCR} at the central blocks of each side road. %
Scenario~B is also split into a case with only \acp{gNB} (macro-only scenario) and another case with \acp{gNB} and \acp{NCR}. %
The system-level performance of these four cases is evaluated through simulations and characterized in terms of \ac{SINR},
\ac{MCS} usage and throughput. %

A simplified version of the Madrid grid~\cite{METIS:D6.1:2013} is considered. %
The scenarios in~\FigRef{NCR_LOS_FIG:scenario} have nine \SI{120}{m}~$\times$~\SI{120}{m} blocks. %
They are surrounded by \SI{3}{m} wide sidewalks and separated from each other by \SI{14}{m} wide streets. 
The \acp{UE} are uniformly positioned on the sidewalks. %
When one of them reaches a block corner, it has \SI{60}{\%} of chance of continuing straight ahead and \SI{20}{\%} of chance to turn left or right. %
These probabilities are proportionally re-scaled whenever some of the directions are not available at a block corner. %

The channel model of~\cite{Pessoa2019}, which is compliant with~\cite{3gpp.38.901c}, is used to model the links between all entities. %
It is consistent in space, time, and frequency dimensions and considers small-scale fading, log-normal shadowing and
distance-dependent path loss propagation components. %
Links from the \acp{gNB} to the \acp{NCR} and to the \acp{UE} follow the \ac{3GPP} \ac{UMa} profile~\cite{3gpp.38.901c}. %
Links from the \acp{NCR} to the \acp{UE} follow the \ac{3GPP} \ac{UMi}~\cite{3gpp.38.901c}. %
From the \acp{gNB} to the \acp{NCR}, we consider \ac{LOS} links, while those from the \acp{gNB} and the \acp{NCR} to the \acp{UE} can be \ac{LOS} or \ac{NLOS}, depending on the \acp{UE}' positions. %
The \acp{NCR} have a fixed amplification gain of \SI{90}{dB} subject to a maximum output power constraint. %
Other system and node parameters are summarized in~\TabRef{NCR_LOS_TAB:Simul_Param}. %

\begin{table}[t]
  \centering
  \scriptsize
  \caption{Simulation parameters.}
  \label{NCR_LOS_TAB:Simul_Param}
  \begin{tabularx}{\columnwidth}{lXXX}
    \toprule
    \textbf{Parameter}                          & \multicolumn{3}{c}{\textbf{Value}}                                        \\
    \midrule
    \multicolumn{4}{l}{\textbf{System}} \\
    \midrule
    Layout                                      & \multicolumn{3}{l}{Simplified Madrid grid~\cite{METIS:D6.1:2013}} \\ 
    Carrier frequency                           & \multicolumn{3}{l}{\SI{28}{\GHz}}                                         \\
    System bandwidth                            & \multicolumn{3}{l}{\SI{50}{\MHz}}                                         \\
    Subcarrier spacing                          & \multicolumn{3}{l}{\SI{60}{\kHz}}                                         \\
    Number of subcarriers per \acs{RB}          & \multicolumn{3}{l}{$12$}                                                  \\
    Number of \acsp{RB}                         & \multicolumn{3}{l}{$66$}                                                  \\
    Slot duration                               & \multicolumn{3}{l}{\SI{0.25}{\ms}}                                        \\
    \acs{OFDM} symbols per slot                 & \multicolumn{3}{l}{$14$}                                                  \\
    Channel generation procedure                & \multicolumn{3}{l}{Cf.~\cite[Fig.~7.6.4-1]{3gpp.38.901c}}                  \\
    Path loss                                   & \multicolumn{3}{l}{Cf.~\cite[Table 7.4.1-1]{3gpp.38.901c}}                 \\
    Fast fading                                 & \multicolumn{3}{l}{Cf.~\cite[Sec.7.5]{3gpp.38.901c} and 
                                                  \cite[Table~7.5-6]{3gpp.38.901c}}                       \\
    \ac{AWGN} power per subcarrier              & \multicolumn{3}{l}{\SI{-174}{dBm}}                                        \\
    Noise figure                                & \multicolumn{3}{l}{\SI{9}{\decibel}}                                      \\
    Mobility model                              & \multicolumn{3}{l}{Pedestrian~\cite{3gpp.37.885b}}                         \\
    Number of \acp{UE}                          & \multicolumn{3}{l}{72}                                                    \\
    \acs{CBR} traffic packet size               & \multicolumn{3}{l}{$3072$ bits}                                           \\
    \acs{CBR} traffic packet inter-arrival time & \multicolumn{3}{l}{\SI{4}{slots}}                                         \\
    \midrule
    \textbf{Nodes}                    & \textbf{\ac{gNB}}               & \textbf{\ac{NCR}}               & \textbf{\acs{UE}} \\
    \midrule
    Height                                      & \SI{25}{\meter}                 & \SI{10}{\meter}                 & \SI{1.5}{\meter} \\
    Transmit power                              & \SI{35}{\decibel m}             & \SI{33}{\decibel m}             & \SI{24}{\decibel m} \\
    Fixed DL/UL amplification gain                              & -             & \SI{90}{\decibel}             & - \\
    Antenna tilt                                & $12^{\circ}$                    & $12^{\circ}$                    & $0^{\circ}$ \\
    Antenna array                               & URA $8\times 8$                 & URA $8\times 8$                 & Single antenna \\
    Antenna element pattern                     & \ac{3GPP} 3D~\cite{3gpp.38.901c} & \ac{3GPP} 3D~\cite{3gpp.38.901c} & Omni \\
    Max. antenna element gain                   & \SI{8}{\decibel i}              & \SI{8}{\decibel i}              & \SI{0}{\decibel i} \\
    Speed                                       & \SI{0}{km/h}                    & \SI{0}{km/h}                    & \SI{3}{km/h} \\
    \bottomrule
  \end{tabularx}
\end{table}

\begin{table*}[]
	\scriptsize
	\caption{\ac{SINR} and throughput values for Scenarios A \& B at the 10$^{\text{th}}$, 50$^{\text{th}}$ and 90$^{\text{th}}$ percentiles.}
	\label{NCR_LOS_TAB:Results-Percentiles}
	\setlength{\extrarowheight}{1pt}
	\centering
	\begin{tabular}{cc|ccccc|ccccc}
		\cline{3-12}
		&
		&
		\multicolumn{5}{c|}{\textbf{Scenario~A}} &
		\multicolumn{5}{c}{\textbf{Scenario~B}} \\ \cline{3-12} 
		&
		&
		\multicolumn{3}{c|}{SINR (dB)} &
		\multicolumn{2}{c|}{Throughput (MBits/s)} &
		\multicolumn{3}{c|}{SINR (dB)} &
		\multicolumn{2}{c}{Throughput (MBits/s)} \\ \hline
		\multicolumn{1}{c|}{\multirow{5}{*}{\rotatebox{90}{Downlink}}} &
		&
		\multicolumn{2}{c|}{Direct links} &
		\multicolumn{1}{c|}{Forwarded links} &
		\multicolumn{2}{c|}{} &
		\multicolumn{2}{c|}{Direct links} &
		\multicolumn{1}{c|}{Forwarded links} &
		\multicolumn{2}{c}{} \\
		\multicolumn{1}{c|}{} &
		Percentile &
		w/o NCRs &
		\multicolumn{1}{c|}{w/ NCRs} &
		\multicolumn{1}{c|}{} &
		w/o NCRs &
		w/ NCRs &
		w/o NCRs &
		\multicolumn{1}{c|}{w/ NCRs} &
		\multicolumn{1}{c|}{} &
		w/o NCRs &
		w/ NCRs \\ \cline{2-12} 
		\multicolumn{1}{c|}{} &
		10$^{\text{th}}$ &
		2.97 &
		\multicolumn{1}{c|}{8.93} &
		\multicolumn{1}{c|}{25.05} &
		1.28 &
		2.54 &
		6.92 &
		\multicolumn{1}{c|}{6.05} &
		\multicolumn{1}{c|}{22.95} &
		3.08 &
		3.20 \\ \cline{2-12} 
		\multicolumn{1}{c|}{} &
		50$^{\text{th}}$ &
		18.13 &
		\multicolumn{1}{c|}{25.72} &
		\multicolumn{1}{c|}{34.94} &
		2.75 &
		3.16 &
		18.79 &
		\multicolumn{1}{c|}{22.05} &
		\multicolumn{1}{c|}{36.81} &
		3.58 &
		4.08 \\ \cline{2-12} 
		\multicolumn{1}{c|}{} &
		90$^{\text{th}}$ &
		46.64 &
		\multicolumn{1}{c|}{47.74} &
		\multicolumn{1}{c|}{47.85} &
		3.40 &
		3.63 &
		35.50 &
		\multicolumn{1}{c|}{43.76} &
		\multicolumn{1}{c|}{48.70} &
		4.50 &
		4.53 \\ \hline
		\multicolumn{1}{c|}{\multirow{5}{*}{\rotatebox{90}{Uplink}}} &
		&
		\multicolumn{2}{c|}{Direct links} &
		\multicolumn{1}{c|}{Forwarded links} &
		&
		&
		\multicolumn{2}{c|}{Direct links} &
		\multicolumn{1}{c|}{Forwarded links} &
		&
		\\
		\multicolumn{1}{c|}{} &
		Percentile &
		w/o NCRs &
		\multicolumn{1}{c|}{w/ NCRs} &
		\multicolumn{1}{c|}{} &
		w/o NCRs &
		w/ NCRs &
		w/o NCRs &
		\multicolumn{1}{c|}{w/ NCRs} &
		\multicolumn{1}{c|}{} &
		w/o NCRs &
		w/ NCRs \\ \cline{2-12} 
		\multicolumn{1}{c|}{} &
		10$^{\text{th}}$ &
		-1.72 &
		\multicolumn{1}{c|}{1.51} &
		\multicolumn{1}{c|}{19.50} &
		0.10 &
		1.81 &
		-0.83 &
		\multicolumn{1}{c|}{1.70} &
		\multicolumn{1}{c|}{21.40} &
		1.87 &
		2.67 \\ \cline{2-12} 
		\multicolumn{1}{c|}{} &
		50$^{\text{th}}$ &
		9.65 &
		\multicolumn{1}{c|}{13.32} &
		\multicolumn{1}{c|}{26.70} &
		2.21 &
		3.01 &
		8.18 &
		\multicolumn{1}{c|}{14.06} &
		\multicolumn{1}{c|}{32.49} &
		3.31 &
		3.69 \\ \cline{2-12} 
		\multicolumn{1}{c|}{} &
		90$^{\text{th}}$ &
		35.66 &
		\multicolumn{1}{c|}{36.04} &
		\multicolumn{1}{c|}{39.52} &
		3.15 &
		3.78 &
		25.46 &
		\multicolumn{1}{c|}{33.06} &
		\multicolumn{1}{c|}{44.15} &
		4.68 &
		5.08 \\ \cline{1-12} 
	\end{tabular}
\end{table*}

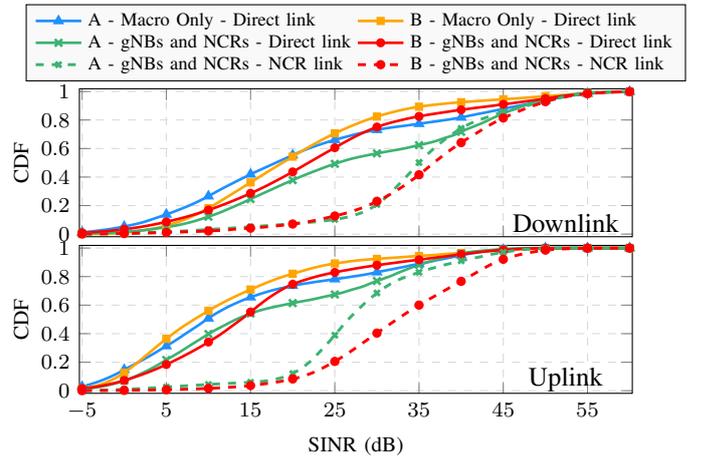
\begin{figure}[t]
	\captionsetup[subfigure]{labelformat=empty}
	
	\subfloat{%
		\begin{tikzpicture}
		\begin{axis}[common plots axis options,
			ylabel=CDF,
			xmin = -5, xmax = 60,
			xtick = {-5,5,...,60},
			xticklabel=\empty,
			legend style={
		    	at = {(0.5, 1.05)},
		    	anchor = south,
		    	legend columns = 2,
			}
			]
			\pgfplotstableread [col sep=comma] {\plotsDataPath/downlink/sinr_all.csv}\tableData
			
			\addplot[scenario01Dir style]
			table[x=Direct_BS_1_x, y=Direct_BS_1_y] from \tableData;
			\addlegendentry{A - Macro Only - Direct link}

			\addplot[scenario02Dir style]
			table[x=Direct_BS_2_x, y=Direct_BS_2_y] from \tableData;
			\addlegendentry{B - Macro Only - Direct link}		
			
			\addplot[scenario311Dir style]
			table[x=Direct_NCR_31_BS_1_x, y=Direct_NCR_31_BS_1_y] from \tableData;
			\addlegendentry{A - gNBs and NCRs - Direct link}
			
			\addplot[scenario322Dir style]
			table[x=Direct_NCR_32_BS_2_x, y=Direct_NCR_32_BS_2_y] from \tableData;
			\addlegendentry{B - gNBs and NCRs - Direct link}	
			
			\addplot[scenario311Ncr style]
			table[x=Ncr_NCR_31_BS_1_x, y=Ncr_NCR_31_BS_1_y] from \tableData;
			\addlegendentry{A - gNBs and NCRs  - NCR link}
			
			\addplot[scenario322Ncr style]
			table[x=Ncr_NCR_32_BS_2_x, y=Ncr_NCR_32_BS_2_y] from \tableData;
			\addlegendentry{B - gNBs and NCRs - NCR link}
		\end{axis}
	\node[at={(6.4, 0.18)}] {Downlink};
	\end{tikzpicture}
	
	\label{NCR_LOS_FIG:Simulation-results-ncr-sinr-all-downlink}
	}
\vspace{-1.85em}
	\subfloat{%
		\begin{tikzpicture}
		\begin{axis}[common plots axis options,
			ylabel=CDF,
			xlabel=SINR (dB),
			xmin = -5, xmax = 60,
			xtick = {-5,5,...,60},
			legend style={
		    	at = {(0.5, 1.05)},
		    	anchor = south,
		    	legend columns = 2,
			}
			]
			\pgfplotstableread [col sep=comma] {\plotsDataPath/uplink/sinr_all.csv}\tableData
			
			\addplot[scenario01Dir style]
			table[x=Direct_BS_1_x, y=Direct_BS_1_y] from \tableData;

			\addplot[scenario02Dir style]
			table[x=Direct_BS_2_x, y=Direct_BS_2_y] from \tableData;
			
			\addplot[scenario311Dir style]
			table[x=Direct_NCR_31_BS_1_x, y=Direct_NCR_31_BS_1_y] from \tableData;
			
			\addplot[scenario322Dir style]
			table[x=Direct_NCR_32_BS_2_x, y=Direct_NCR_32_BS_2_y] from \tableData;
			
			\addplot[scenario311Ncr style]
			table[x=Ncr_NCR_31_BS_1_x, y=Ncr_NCR_31_BS_1_y] from \tableData;
			
			\addplot[scenario322Ncr style]
			table[x=Ncr_NCR_32_BS_2_x, y=Ncr_NCR_32_BS_2_y] from \tableData;
		\end{axis}
	\node[at={(6.4, 0.18)}] {Uplink};
	\end{tikzpicture}
	
	\label{NCR_LOS_FIG:Simulation-results-ncr-sinr-all-uplink}
	}

\caption{\acs{CDF} of \acs{SINR} for the direct and forwarded link for Scenarios A and B on Downlink and Uplink, respectively.}
\label{NCR_LOS_FIG:Simulation-results-sinr-cdf-all}
\end{figure}

Figure~\ref{NCR_LOS_FIG:Simulation-results-sinr-cdf-all} presents the \acp{CDF} of the \acp{SINR} of the direct and forwarded links for both Scenarios A and B. %
Solid curves refer to direct links between \acp{gNB} and \acp{UE}, while dashed curves refer to links forwarded by \acp{NCR}. %
The blue and green curves are obtained for Scenario~A without and with \acp{NCR}, respectively, while the yellow and red ones relate to Scenario~B analogously. %

Considering the \ac{DL}, at the 10$^{\text{th}}$ percentile of \ac{SINR} \acp{CDF} shown in~\FigRef{NCR_LOS_FIG:Simulation-results-sinr-cdf-all} and summarized in~\TabRef{NCR_LOS_TAB:Results-Percentiles}, Scenario~A achieved \ac{SINR} of \SI{2.97}{dB} for the direct links in the macro-only case. %
In Scenario~A with \acp{NCR}, at the 10$^\text{th}$ percentile, we observe an \ac{SINR} of \SI{8.93}{dB} for the direct links and \SI{25.05}{dB} for the forwarded links, respectively. %
Therefore, compared to the cases with macro-only, the worst connections are improved by around \SI{6}{dB} when connected to the \acp{gNB} and by \SI{22}{dB} when served through the \acp{NCR}. %
This is intuitive because the introduction of the \acp{NCR} reduces the effective distance between \ac{UE} and their serving nodes. %
Regarding the 50$^{\text{th}}$ percentile of the \ac{SINR} in the same scenarios, the value of \SI{18.13}{dB} for the macro-only case improves to \SI{25.72}{dB} for the direct links and to \SI{34.94}{dB} for the \ac{NCR}-forwarded links, achieving gains of $\approx$ \SI{7.60}{dB} and \SI{16.81}{dB}, respectively. %
With respect to the 90$^{\text{th}}$ percentile of the \ac{SINR} in the same scenarios, the value of \SI{46.64}{dB} for the macro-only case improves to \SI{47.74}{dB} for the direct links and to \SI{47.85}{dB} for the \ac{NCR}-forwarded links, i.e., we have gains of $\approx$ \SI{1}{dB} and \SI{1.20}{dB}, respectively. %

As for the \ac{UL}, Scenario~A at the 10$^{\text{th}}$ percentile of \ac{SINR} \acp{CDF}, also shown in~\FigRef{NCR_LOS_FIG:Simulation-results-sinr-cdf-all} and summarized in~\TabRef{NCR_LOS_TAB:Results-Percentiles}, had an \ac{SINR} of \SI{-1.72}{dB} for the direct links considering the deployment composed only of \acp{gNB}, while having \SI{1.51}{dB} and \SI{19.50}{dB} for the direct and forwarded links when using \acp{NCR}. %
Therefore, by using \acp{NCR}, the worst connections are improved by around \SI{3.23}{dB} when connected to the \acp{gNB} and by \SI{21.22}{dB} when served through the \acp{NCR}. %
For the 50$^{\text{th}}$ percentile of the \ac{SINR}, the value of \SI{9.65}{dB} for the macro-only case improves to \SI{13.32}{dB} for the direct links and \SI{26.70}{dB} for the forwarded links, achieving gains of $\approx$ \SI{3.70}{dB} and \SI{17.05}{dB}, respectively. %
Furthermore, for the 90$^{\text{th}}$ percentile of the \ac{SINR}, the value of \SI{35.66}{dB} for the macro-only case improves to \SI{36.04}{dB} for the direct links and \SI{39.52}{dB} for the forwarded links, i.e., we have gains of $\approx$ \SI{0.38}{dB} and \SI{3.86}{dB}, respectively. %

For Scenario~B \ac{DL}, we have similar trends in terms of \ac{SINR} at the 10$^{\text{th}}$ percentile with a value of \SI{6.92}{dB} for the direct links in the macro-only deployment, while for the case with \acp{NCR} we observe \SI{6.05}{dB} for direct links and \SI{22.95}{dB} for the forwarded links. %
For the 50$^{\text{th}}$ percentile of the \ac{SINR}, the value of \SI{18.79}{dB} for the macro-only case improves to \SI{22.05}{dB} for the direct links and \SI{36.81}{dB} for the forwarded links, achieving gains of $\approx$ \SI{3.26}{dB} and \SI{18.02}{dB}, respectively. %
In Scenario~B, it is perceived a more expressive relative growth at the 90$^{\text{th}}$ percentile of the \ac{SINR} values between the cases without and with \ac{NCR}. %
We observe \SI{35.50}{dB} for direct links in the macro-only case and \SI{43.76}{dB} and \SI{48.70}{dB} for the direct and forwarded links \ac{DL}, respectively, in the case with \acp{NCR}. %
The same trend is observed for the cases with \ac{UL} communication. %

While the absolute values of \ac{SINR} in Scenario~B are generally lower than those seen in Scenario~A, the relative improvements are higher. %
This is because Scenario~B has four macros which introduce more interference in the system and lead to lower overall \ac{SINR} values. %
In this more challenging interference scenario, the relative gains that can be brought by \ac{NCR} are naturally larger. %
Notice that for the forwarded links, which remain relatively separated from each other in both Scenarios A and B, the \acp{NCR}' performance remains similar to each other. %
In this way, the presence of \acp{NCR} improves the \ac{QoS} experienced by the weak \acp{UE} significantly, in harmony with the main objectives considered by \ac{3GPP} Rel-18. %
However, to limit the interference, one needs to determine the proper number of \acp{NCR} and deploy them with proper planning. %
Moreover, due to the lower transmit power of the \acp{UE}, compared to \acp{gNB}, the \ac{UL} direction is the one that benefits the most from the deployment of \acp{NCR}. %

It is interesting to note that we have observed the same qualitative conclusions when considering different \acp{MCS} (although not demonstrated in the figures). %
For the highest \ac{MCS}, i.e., \ac{MCS}~15, we have observed a 40\% usage for Scenario~A without \acp{NCR} and 70\% for the Scenario~A with \acp{NCR} while for Scenario~B, we have observed a 30\% usage with macro-only and 50\% with \acp{NCR}. %
In absolute terms, Scenario~A has in general larger usage of the highest \ac{MCS} than Scenario~B, while the relative
increase (fourfold) for the highest \ac{MCS} usage in Scenario~B is larger. %

\begin{figure}[t]
\centering
	\begin{tikzpicture}
		\begin{axis}[common plots axis options,
			ylabel=CDF,
			xlabel=Throughput (Mbps),
                        xmin=-0.001, xmax = 5.12,
                        ymin=-0.001, ymax=1.01,
			xtick = {0,0.5,...,4.5},
			width=9.1cm,
			height=4.5cm,
			legend style={
		    	at = {(0.5, 1.05)},
		    	anchor = south,
		    	legend columns = 2,
			}
			]
			\pgfplotstableread [col sep=comma] {\plotsDataPath/downlink/throughput_all.csv}\tableDataDL
			
			\pgfplotstableread [col sep=comma] {\plotsDataPath/uplink/throughput_all.csv}\tableDataUL

			\addplot[scenario01Dir style]
			table[x=BS_1_x, y=BS_1_y] from \tableDataDL;
			\addlegendentry{DL-A - Macro only}
			
			\addplot[scenario02Dir style]
			table[x=BS_2_x, y=BS_2_y] from \tableDataDL;
			\addlegendentry{DL-B - Macro only}
			
			\addplot[scenario311Dir style]
			table[x=NCR_31_BS_1_x, y=NCR_31_BS_1_y] from \tableDataDL;
			\addlegendentry{DL-A - gNBs and NCRs}
			
			\addplot[scenario322Dir style]
			table[x=NCR_32_BS_2_x, y=NCR_32_BS_2_y] from \tableDataDL;
			\addlegendentry{DL-B - gNBs and NCRs}
			\addplot[scenario01Ncr style]
			table[x=BS_1_x, y=BS_1_y] from \tableDataUL;
			\addlegendentry{UL-A- Macro only}
			
			\addplot[scenario02Ncr style]
			table[x=BS_2_x, y=BS_2_y] from \tableDataUL;
			\addlegendentry{UL-B - Macro only}
			
			\addplot[scenario311Ncr style]
			table[x=NCR_31_BS_1_x, y=NCR_31_BS_1_y] from \tableDataUL;
			\addlegendentry{UL-A - gNBs and NCRs}
			
			\addplot[scenario322Ncr style]
			table[x=NCR_32_BS_2_x, y=NCR_32_BS_2_y] from \tableDataUL;
			\addlegendentry{UL-B - gNBs and NCRs}
		\end{axis}
	\end{tikzpicture}
\label{NCR_LOS_FIG:Simulation-results-ncr_throughput_tot}
\caption{\ac{CDF} of throughput.}
\label{NCR_LOS_FIG:throughput-cdf}
\end{figure}
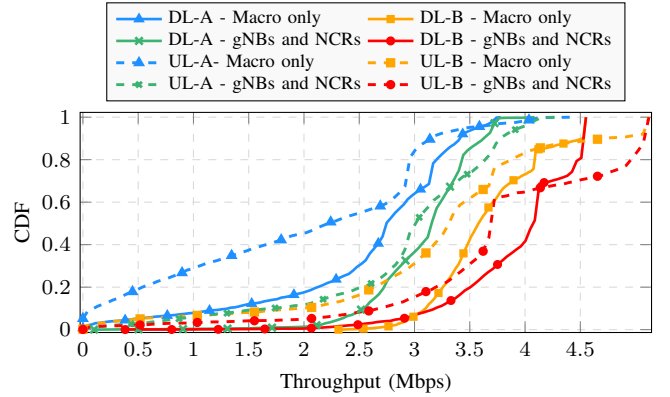

Figure~\ref{NCR_LOS_FIG:throughput-cdf} shows the \acp{CDF} of the throughput values for Scenarios A and B. %
Solid and dashed lines refer to \ac{DL} and \ac{UL}, respectively. %
The blue and green curves refer to Scenario~A, without and with \acp{NCR}, respectively, while, analogously, the yellow and red curves refer to Scenario~B. %
As demonstrated in~\FigRef{NCR_LOS_FIG:throughput-cdf} and~\TabRef{NCR_LOS_TAB:Results-Percentiles},
for Scenario~A in \ac{UL}, the 10$^\text{th}$ percentile throughput is \SI{0.10}{MBits/s} and \SI{1.81}{MBits/s} without and with \acp{NCR}, respectively. The 50$^\text{th}$ percentile throughput is \SI{2.21}{MBits/s} and \SI{3.01}{MBits/s} without and with \acp{NCR}, respectively, while the 90$^\text{th}$ percentile throughput is \SI{3.15}{MBits/s} and \SI{3.78}{MBits/s}, respectively. %
That means an eighteen-fold improvement for the worst links in \ac{UL} when \acp{NCR} are used in Scenario~A. %
That is, the \acp{UE} with poor direct link to the \ac{gNB} (for instance, cell-edge \acp{UE}) benefit most from the presence of \acp{NCR}, although even the cell-center \acp{UE} still benefit slightly from the presence of the \acp{NCR}. %

For Scenario~B in \ac{UL}, the 10$^\text{th}$ percentile throughput is \SI{1.87}{MBits/s} and \SI{2.67}{MBits/s} without
and with \acp{NCR}, respectively. The 50$^\text{th}$ percentile throughput is \SI{3.31}{MBits/s} and \SI{3.69}{MBits/s} without
and with \acp{NCR}, respectively, while the 90$^\text{th}$ percentile throughput is \SI{4.68}{MBits/s} and \SI{5.08}{MBits/s}, respectively. %
The different values of throughput gain between Scenarios A and B are explained by the higher interference level in the latter which strongly impacts the worst links. %
On the other hand, since in Scenario~B without and with \acp{NCR} we have larger number of nodes reusing the spectrum, the overall throughput obtained in Scenario~B in both \ac{DL} and \ac{UL} is larger in absolute values, compared to those obtained in Scenario~A (see~\FigRef{NCR_LOS_FIG:throughput-cdf}). %
Note that very high but different \ac{SINR} values may lead to the same throughput, since this latter is ultimately limited by the capacity of the highest \ac{MCS}.

\FloatBarrier
\section{On the Challenges of \acs{NCR}-assisted Networks}
\label{NCR_LOS_SEC:Challenges-of-NCR-assisted-networks}
Although the system-level evaluations show great potential for coverage improvement via the implementation of \acp{NCR}, there are still challenges to be considered before the \ac{NCR} can be implemented on a large-scale. %
Some of them are listed in the following:
\begin{itemize}
	\item Lack of testbed evaluations: To the best of our knowledge, there is yet no practical implementation of \acp{NCR}. %
	Once \acp{NCR} are manufactured, testbed evaluations and field measurements can provide better understanding of the potentials and challenges of \acp{NCR}. %
	
	\item (Self-)Interference management: Rel-17 repeaters support up to \SI{90}{dB} amplification gains and \acp{NCR} are expected to support at least the same number. %
	In such cases, self-interference as well as interference to/from the other nodes can be an issue which requires accurate interference management mechanisms. %
	Here, network planning and beamforming are expected to reduce the effect of (self-)interference remarkably. %
	However, in scenarios where the spatial deployment is not able to filter (self-)interference links, (self-)interference effects need to be considered. %
	
	\item Cost-efficiency tradeoff: Although \ac{NCR} can improve the network coverage, its performance, in terms of both cost and efficiency, needs to be further evaluated compared to its alternative competitors such as \ac{IAB} and \ac{RIS}. %
	Particularly, while \ac{IAB} is a more complex node than \ac{NCR}, it covers a larger area. %
	Thus, compared to \acp{NCR}, a fewer number of \ac{IAB} nodes may be required to cover an area, and the total cost of ownership, on the network level, for these technologies need to be carefully investigated. %
	On the other hand, \ac{NCR} is expected to outperform the \ac{RIS} in terms of performance, with limited or no cost increment (see~\cite{Guo2022} for a conceptual comparison between the \ac{RIS} and \ac{NCR}). %
\end{itemize}

\section{Conclusions and Future Perspectives}
\label{NCR_LOS_SEC:Conclusion}

In this paper, we introduced the \ac{NCR} concept, the main standardization effort on it, as well as its potential and challenges in different deployments. %
As shown, employing \acp{NCR} improves the performance of both direct and forwarded links considering different performance metrics. %
Particularly, cell-edge \acp{UE} and \ac{UL} communications are the ones with the highest benefit from the presence of \acp{NCR} in \ac{mmWave} networks. %
Furthermore, the results show the importance of network planning where the relative performance gains achievable using \ac{NCR} depend on the number and positions of these nodes selected for deployment. %
Depending on the number of nodes, resource reuse and interference levels, the introduction of \acp{NCR} may result in \ac{UL} throughput gain of up to ten times for the cell-edge \acp{UE}. %

\section*{Acknowledgments}

This work was supported by Ericsson Research, Sweden, and Ericsson Innovation Center, Brazil, under UFC.51 Technical Cooperation Contract Ericsson/UFC. The work of Victor F. Monteiro was supported by CNPq under Grant 308267/2022-2. The work of Tarcisio F. Maciel was supported by CNPq under Grant 312471/2021-1. The work of Francisco R. M. Lima was supported by FUNCAP (edital BPI) under Grant BP4-0172-00245.01.00/20.


\printbibliography{}

@STRING{IEEE_J_AP         = "{IEEE} Trans. Antennas Propag."}

@STRING{IEEE_O_ACC        = "{IEEE} Access"}

@String { IEEE_J_OCS                = {{IEEE} {O}pen {J}. {C}ommu. {S}oc.} }

@String { WIOPT                     = {{P}roc. of the {I}nternat. {S}ymp. on {M}odeling and {O}pt. in {M}obile, {A}d {H}oc, and {W}ireless {N}etw. {(WiOPT)}} }

@String { std3GPP          = {3rd Generation Partnership Project {(3GPP)}} }

@TechReport{3gpp.38.867,
	Title                    = {{Study on NR network-controlled repeaters}},
	Author                   = {3GPP},
	Institution              = std3GPP,
	Year                     = {2022},
	Month                    = sep,
	Number                   = {{38.867}},
	Type                     = {TR},
	Note					 = {v.18.0.0},
	Url                      = {http://www.3gpp.org/DynaReport/38867.htm}
}

@TechReport{3gpp.37.885b,
	Title                    = {Study on evaluation methodology of new Vehicle-to-Everything {(V2X)} use cases for {LTE} and {NR}},
	Author                   = {3GPP},
	Institution              = std3GPP,
	Year                     = {2019},
	Month                    = jun,
	Number                   = {{37.885}},
	Type                     = {TS},
	Note                     = {v.15.3.0},
	Url                      = {http://www.3gpp.org/ftp/Specs/html-info/37885.htm},
	Urldate		   		     = {2019-04-17}
}

@TechReport{3gpp.38.331b,
	Title                    = {{NR}; Radio resource control {(RRC)} protocol specification},
	Author                   = {3GPP},
	Institution              = std3GPP,
	Year                     = {2023},
	Month                    = sep,
	Number                   = {38.331},
	Type                     = {TS},
	Version                  = {17.6.0},
	Url                      = {http://www.3gpp.org/ftp/Specs/html-info/38331.htm},
	Urldate			         = {2018-10-01}
}

@TechReport{3gpp.38.306,
  Title                    = {{5G;NR;User Equipment (UE) radio access capabilities }},
  Author                   = {3GPP},
  Institution              = std3GPP,
  Year                     = {2023},
  Month                    = oct,
  Number                   = {{38.306}},
  Type                     = {TS},
  Note                     = {v.17.6.0},
  Url                      = {https://www.3gpp.org/ftp/Specs/archive/38_series/38.306/},
  Urldate		   		   = {2023-09-28}
}

@TechReport{3gpp.38.901c,
	Title                    = {Study on Channel Model for Frequencies from 0.5 to 100 {GHz}},
	Author                   = {3GPP},
	Institution              = std3GPP,
	Year                     = {2022},
	Month                    = mar,
	Number                   = {{38.901}},
	Type                     = {TR},
	Note                     = {v.17.0.0},
	Url                      = {http://www.3gpp.org/DynaReport/38901.htm},
	Urldate		   		   = {2017-09-26}
}

@TechReport{3gpp.38.174c,
	Title                    = {{NR}; Integrated access and backhaul radio transmission and Reception},
	Author                   = {3GPP},
	Institution              = std3GPP,
	Year                     = {2023},
	Month                    = sep,
	Number                   = {{38.174}},
	Type                     = {TS},
	Note                     = {v.18.2.0},
	Url                      = {http://www.3gpp.org/ftp/Specs/html-info/38174.htm},
	Urldate		   	 = {2023-06-16}
}

@TechReport{3gpp.38.106,
	Title                    = {{NR}; {NR} Repeater Radio Transmission and Reception},
	Author                   = {3GPP},
	Institution              = std3GPP,
	Year                     = {2023},
	Month                    = mar,
	Number                   = {{38.106}},
	Type                     = {TS},
	Note                     = {v.17.4.0},
	Url                      = {http://www.3gpp.org/ftp/Specs/html-info/38106.htm},
  	Urldate		   	 = {2023-06-16}
}

@TechReport{3gpp.RP-222673,
  Title                    = {{New WID on NR network-controlled repeaters}},
  Author                   = {3GPP},
  Institution              = std3GPP,
  Year                     = {2022},
  Month                    = sep,
  Number                   = {{222673}},
  Type                     = {RP},
  Note                     = {TSG RAN meeting no. 97-e},
  Url                      = {https://www.3gpp.org/ftp/tsg_ran/TSG_RAN/TSGR_97e/Docs/RP-222673.zip},
  Urldate		   = {2023-06-16}
}

@Article{Pessoa2019,
  author =       {Alexandre Matos Pessoa and Igor Moaco Guerreiro and Carlos Filipe Moreira e Silva and Tarcisio Ferreira Maciel and Diego Aguiar Sousa and Darlan Cavalcante Moreira and Francisco Rodrigo Porto Cavalcanti},
  title =        {A Stochastic Channel Model With Dual Mobility for {5G} Massive Networks},
  journaltitle = IEEE_O_ACC,
  year =         2019,
  volume =       7,
  month =        oct,
  pages =        {149971--149987},
  issn =         {2169-3536},
  doi =          {10.1109/ACCESS.2019.2947407},
}

@Article{Madapatha2020,
  author =       {Charitha Madapatha and Behrooz Makki and Chao Fang and Oumer Teyeb and Erik Dahlman and Mohamed-Slim Alouini and Tommy Svensson},
  title =        {On Integrated Access and Backhaul Networks: Current Status and Potentials},
  journaltitle = IEEE_J_OCS,
  year =         2020,
  volume =       1,
  month =        sep,
  pages =        {1374--1389},
  doi =          {10.1109/OJCOMS.2020.3022529},
}

@Report{METIS:D6.1:2013,
  author =       {Patrick Agyapong and others},
  title =        {Simulation guidelines},
  type =         {Deliverable},
  institution =  {METIS},
  year =         2013,
  number =       {6.1},
  month =        oct,
  url =          {https://metis2020.com/wp-content/uploads/deliverables/METIS_D6.1_v1.pdf},
  urldate =      {2021-09-20},
}

@Online{Guo2022,
	author =       {Hao Guo and Charitha Madapatha and Behrooz Makki and Boris Dortschy and Lei Bao and Magnus Åström and Tommy Svensson},
	title =        {A Comparison between Network-Controlled Repeaters and Reconfigurable Intelligent Surfaces},
	year =         2022,
	eprint=		   {2211.06974},
	month =        nov,
    primaryclass = {cs.NI},
	eprinttype =   {arXiv},
}

@Article{Flamini2022,
	author =       {Flamini, Roberto and De Donno, Danilo and Gambini, Jonathan and Giuppi, Francesco and Mazzucco, Christian and Milani, Angelo and Resteghini, Laura},
	title =        {Towards a Heterogeneous Smart Electromagnetic Environment for Millimeter-Wave Communications: An Industrial Viewpoint},
	journaltitle = IEEE_J_AP,
	year =         2022,
	volume =       70,
	number =       10,
	month =        feb,
	pages =        {8898--8910},
	doi =          {10.1109/TAP.2022.3151978},
}

@InProceedings{Leone2022,
  author =       {Giuseppe Leone and Eugenio Moro and Ilario Filippini and Antonio Capone and Danilo De Donno},
  title =        {Towards Reliable {mmWave} {6G} {RAN}: Reconfigurable Surfaces, Smart Repeaters, or Both?},
  year =         2022,
  booktitle =    WIOPT,
  month =        sep,
  pages =        {1--9},
  doi =          {10.1109/AERO.2015.7118906},
  address = 	     {Turin, Italy}
}

@Article{Monteiro2022,
  author =       {Victor Farias Monteiro and Fco. Rafael Marques Lima and Darlan Cavalcante Moreira and Diego Aguiar Sousa and Tarcisio Ferreira Maciel and Behrooz Makki and Hans Hannu},
  title =        {Paving the Way Towards Mobile {IAB}: Problems, Solutions and Challenges},
  journal =      IEEE_J_OCS,
  year =         2022,
  volume =       3,
  month =        nov,
  pages =        {2347-2379},
  doi =          {10.1109/OJCOMS.2022.3224576},
}
\endrefsection{}
\acbarrier{}

\section*{Biographies}
{\footnotesize
	\begin{itemize}
		
		\item \textbf{Fco. Italo G. Carvalho} (italoguedes@gtel.ufc.br) is an undergraduate student at the Federal University of Cear\'{a} (UFC), Fortaleza, Brazil and is with the Wireless Telecommunications Research Group (GTEL), UFC. 
		
		\item \textbf{Raul Victor de O. Paiva} (raul.paiva@gtel.ufc.br) is a PhD student at UFC and is with GTEL.
		
		\item \textbf{Tarcisio F. Maciel} (maciel@gtel.ufc.br) is a Professor at UFC and is with the GTEL. He is interested on multiuser/multiantenna communications.
		
		\item \textbf{Victor F. Monteiro} (victor@gtel.ufc.br) is a Professor at UFC and is with the GTEL. He is focused on network architecture and protocols.
		
		\item \textbf{Fco. Rafael M. Lima} (rafaelm@gtel.ufc.br) is a Professor at UFC and is with the GTEL. He  has focused on radio resource management.
		
		\item \textbf{Darlan C. Moreira} (darlan@gtel.ufc.br) is a senior researcher at GTEL focusing on simulation programming. 
		
		\item \textbf{Diego A. Sousa} (diego@gtel.ufc.br) is a Professor at the Federal Institute of Education, Science, and Technology of Cear\'{a} (IFCE), Paracuru, Brazil and is with the GTEL. He has focused on radio resource management.
	
		\item \textbf{Behrooz Makki} (behrooz.makki@ericsson.com) is a Senior Researcher in Ericsson, focusing on different aspects of network densification.
				
		\item \textbf{Magnus Åstr$\ddot{\text{o}}$m} (magnus.astrom@ericsson.com) is a Research Leader in Ericsson, developing different aspects of 5G and beyond.
		
		\item \textbf{Lei Bao} (lei.bao@ericsson.com) is a Senior Researcher in Ericsson, focusing on wireless backhaul.

\end{itemize}}


\ifCLASSOPTIONcaptionsoff
  \newpage
\fi

\end{document}